Biological Sciences

Biophysics, Evolution

# The Emergence of Scaling in Sequence-based Physical Models of Protein Evolution

Eric J. Deeds* and Eugene I. Shakhnovich†‡

*Department of Molecular and Cellular Biology, Harvard University, 7 Divinity Avenue, Cambridge, MA 02138*

*†Department of Chemistry and Chemical Biology, Harvard University, 12 Oxford Street, Cambridge, MA 02138*

‡Corresponding Author: Eugene Shakhnovich
Department of Chemistry and Chemical Biology
Harvard University
12 Oxford Street
Cambridge, MA 02138
Tel: 617.495.4130
Fax: 617.384.9228
Email: eugene@vodka.chem.harvard.edu

**It has recently been discovered that many biological systems, when represented as graphs, exhibit a *scale-free* topology. One such system is the set of structural relationships among protein domains. The scale-free nature of this and other systems has previously been explained using network growth models that, while motivated by biological processes, do not explicitly consider the underlying physics or biology. In the present work we explore a sequence-based model for the evolution protein structures and demonstrate that this model is able to recapitulate the scale-free nature observed in graphs of real protein structures. We find that this model also reproduces other statistical feature of the protein domain graph. This represents, to our knowledge, the first such microscopic, physics-based evolutionary model for a scale-free network of biological importance and as such has strong implications for our understanding of the evolution of protein structures and of other biological networks.**

Protein structural evolution, and specifically the discovery of new sequence-structure pairs, represents one of the most important facets of molecular evolution (1). Recently, our understanding of structural evolution has advanced considerably, based at least in part on the application of graph theoretic methods to the study of protein structural similarity (1-4). One such application is the Protein Domain Universe Graph (PDUG), which is constructed by representing the non-redundant set of protein structural domains as nodes and using the structural similarity between those domains to define the edges on the graph (2). Analysis of the PDUG demonstrated that the distribution of the number of structural neighbors $k$ per domain (known as the degree distribution, or $p(k)$) follows a power law; that is, $p(k) \sim k^{-\gamma}$ where $\gamma \sim 1.6$, a finding that indicates that the PDUG is a *scale-free* network (2, 5). This observation, along with other features of the PDUG and proteome-specific subgraphs, has lead to the conclusion that structural evolution has been largely divergent in nature, with existing sequence-structure pairs giving rise to new structures through processes such as duplication and divergence. One of the major pieces of evidence for this divergent paradigm has been the observation that graph evolution models based on "divergent" rules are able to create graphs with power-law degree distributions that have exponents $\gamma \sim 1.6$ (2, 4). These models, like most models of the evolution of biological scale-free networks (5-7), are entirely arbitrary; that is, although they attempt to mimic mechanisms such as duplication and divergence, they do not directly model those processes. Thus a major outstanding question in structural evolution revolves around whether or not models based on the *a priori* evolution of actual protein sequences could result in scale-free networks similar to that of the PDUG.

One of the major obstacles to building such a model is the fact that the protein folding problem remains unsolved for structures with realistic degrees of freedom (8), making it difficult to accurately model the evolution of actual polypeptides. Model systems exist, however, in which the folding problem has been solved, and it is possible to approach the question of sequence evolution in such systems. Lattice polymers are one such system, and extensive study of such polymers has provided insight into protein folding, designability and protein evolution (9-16). Although lattice polymers are indeed only a crude approximation to real protein structures, the fact that lattice sequences can posses and fold into unique native structures captures one of the key features of real proteins. In this work, we focus on maximally compact 27-mers on the 3x3x3 cubic lattice. The 27-mer represents a particularly interesting lattice system due to the fact that all maximally compact conformations of this polymer may be enumerated (9), and recent studies have revealed that a graph based on the structural similarity between all of these possible structures (constructed in a manner similar to that used to make the PDUG) exhibits a degree distribution similar to a random graph (15). Furthermore, it has been demonstrated that subgraphs of this Lattice Structure Graph (LSG) can exhibit scale-free degree distributions when the structures are sampled according to divergent evolutionary rules (15).

Although the existence of scale-free "evolved" subgraphs of the LSG is suggestive, these graphs are obtained using algorithms that evaluate the structural similarity of "candidate" nodes to existing nodes in order to determine if they will indeed be added to the evolving graph (15). Such calculations are most likely not performed by organisms as they evolve, and thus the question thus remains as to whether sequence dynamics alone can explain the emergence of scale-free networks from the entire set of possible lattice structures. In the present study we demonstrate that models based solely on the duplication, divergence and folding of lattice sequences can also result in model graphs that are similar to the PDUG. We demonstrate this similarity not only in terms of the traditional degree distribution but also in terms of other statistical features of these graphs. To our knowledge this represents the first instance in which a model based solely on physical and biological mechanisms has reproduced the features of a biologically relevant scale-free network.

**Methods**

**Lattice Model** As discussed above, our evolutionary model is based on the standard physics of lattice polymers (9, 10, 12, 13, 16, 17). The potential energy of a sequence in a given lattice conformation is based on the contacts between monomers that occur in that structure, i.e.:

$$E_c = \sum_{i=1}^{L} \sum_{j=i+1}^{L} \epsilon_{s_i s_j} \Delta_{ij}$$

where $E_c$ is the potential energy of the sequence in conformation $c$, $L$ is the length of the polymer (in this case 27), $\varepsilon_{s_i s_j}$ is the potential energy of a contact between beads of type $s_i$ and $s_j$ in the sequence and $\Delta_{ij}$ is set to 1 if positions $i$ and $j$ are in contact in conformation $c$ and 0 otherwise. Residues are defined to be in contact if they are neighbors in space but not neighbors in sequence. The matrix of contact energies is taken from the Mirny-Shakhnovich (MS) potential and is very similar to the potential of Miyazawa and Jernigan (10, 18, 19). Folding in this model may be assayed using a Z-score technique (10, 12, 17, 20); the Z-score of the native state is defined as:

$$Z_{nat} = \frac{E_{nat} - \langle E \rangle}{\sigma_E}$$

where $E_{nat}$ is the energy of the native state, $\langle E \rangle$ is the average energy of the sequence in all 103346 compact conformations and $\sigma_E$ is the standard deviation in energy for the entire compact ensemble. When the native state is much more stable than the average member of the non-native ensemble, i.e. when the Z-score for the native state has a large negative value, the sequence is assumed to fold into the native state. This method allows for fast evaluation of the folding of sequences and has been used successfully in other contexts (10, 12).

**Evolutionary Algorithm** Our evolutionary model is built on the above physical model and represents a simple interpretation of duplication and divergence. The algorithm begins with a single sequence that has been designed to fold into an arbitrarily chosen lattice structure with a Z-score of less than –7. In each case, folding of the seed sequence into the seed structure is verified using standard Monte Carlo lattice folding techniques (10-12, 16). At each step of the algorithm, an existing sequence-structure pair is randomly chosen for duplication. One of the duplicate sequences is then subjected to a number of mutations (*m*). The algorithm then identifies the new native state of this modified sequence by determining the lowest energy conformation out of all compact possibilities. The Z-score of the newly evolved sequence in this native structure is then checked to determine if the sequence will fold according to some Z-score cutoff. If the sequence folds, the newly evolved sequence-structure pair is added to the model graph; if not, the sequence is discarded and a new sequence is randomly chosen for duplication. The features of this model are diagrammed in Fig. 1. In all cases we evolve 3500 structures using our algorithm in order to obtain graphs with a number of nodes similar to the number of nodes in the PDUG.

**Constructing Graphs of Lattice Structures** Structural similarity between conformations on the lattice is defined according to (15); this method is based on calculating the statistical significance (S-score) of the overlap between two contact maps defining two conformations. The nodes in each graph are constructed from the set of lattice structures chosen by the evolutionary algorithm. Edges are drawn at structural similarity cutoffs ($S_{min}$'s) chosen according to the transition in the giant component of each graph (2, 5, 15). The $S_{min}$ chosen according to this method is between 7 and 8 for all of our evolved graphs, a cutoff that is very similar to the cutoff found for graphs evolved according to our earlier lattice-based model (15).

## Results

**Sequence Evolution without a Folding Constraint**

In order to determine whether sequence dynamics of the simplest kind could reproduce scale-free networks with $\gamma \sim 1.6$, the first runs of this model are performed without a folding cutoff: in this case, every new native state is added to the graph regardless of its ability to fold (this is equivalent to setting the Z-score cutoff in Fig. 1 to positive infinity). Although this instantiation of the model does not implement the restraint of protein folding, it is important to note that the choice of each new structure as the conformation in which the new sequence exhibits the lowest energy represents an important "physical" component of the algorithm. When $m$, the number of mutations per duplication step, is set to 1, the resulting graphs do indeed exhibit scale-free degree distributions, but in these cases we find values of $\gamma$ around 1 for most runs of the algorithm (Fig. 2 *A*), indicating that point mutations alone are insufficient to produce PDUG-like behavior even in the absence of folding constraints. When $m$ is increased to 2, however, networks with $\gamma \sim 1.6$ are readily observed (Fig. 2B), while $m = 3$ results in scale-free networks with $\gamma \sim 2$ (Fig. 2C). For this particular model, various runs with the same parameters yield similar graphs: in the case of setting $m$ to 2, the graphs that are evolved exhibit exponents in the range of 1.4 to 1.6. Indeed, if a second arbitrarily chosen (but structurally unrelated) sequence-structure pair is used to seed the algorithm, graphs with exponents of ~1.6 (with a similar range) are observed for $m = 2$ (for a representative run see Fig. 2D).

These results indicate that "PDUG-like" scale-free networks may be sampled from the underlying random-graph topology of the LSG solely based on the divergence of sequences into new native states. Furthermore, for the above model we find that the amount of sequence divergence employed by the model is intimately related to the observed exponent in the evolved graph. In a certain sense this power-law exponent represents a gross measure of the structural diversity of the graph (i.e. the number of orphan and sparsely connected nodes as compared to the number of highly connected nodes). Thus the dependence of $\gamma$ on $m$ is relatively intuitive: the greater the level of sequence divergence employed in the model, the greater the level of structural diversity one observes in the evolved graph.

**Sequence Evolution with a Folding Criterion**

Real proteins, of course, are subject to rather stringent folding criteria, and so a more realistic set of runs of the model were performed with a folding Z-score cutoff. For the purposes of this work, the cutoff is set to −6 in a heuristic manner: we find that the model runs prohibitively slowly when significantly more stringent folding criteria are applied. Sequences evolved at this cutoff do, however, reliably fold to their native states in Monte Carlo simulations (we tested this for a set of 10 randomly chosen sequence structure pairs evolved under this cutoff, data not shown); sequences evolved under less stringent criteria do not fold as reliably (also, see (12)). Given this folding criterion, we find that the model requires a much larger value of *m* to obtain graphs with exponents of 1.6. For one particular starting structure, we find that *m* = 2 (which gave PDUG-like behavior in the non-folding model above) leads to graphs with exponents ~ 1 (see Fig. 3A). This result indicates that, when a folding constraint is imposed, the algorithm tends to select structures that are highly similar to the original structure when the number of mutations is small, leading to graphs that lack the structural diversity characteristic of the PDUG. Indeed, graphs with exponents similar to that of the PDUG are only readily observed from this starting structure when *m* is set to 8 (see Fig. 3B). It is important to note that this result does not imply that real proteins evolve on the basis of large numbers of simultaneous mutations: it simply indicates that a large amount of sequence divergence is necessary to observe PDUG-like behavior in this lattice model.

The number of mutations required to observe an exponent of 1.6 depends strongly on the starting structure. As mentioned above, an *m* of 8 is sufficient to observe PDUG-like graphs for a given starting sequence and structure. For the alternate seed sequence-structure pair discussed above, however, setting *m* to 8 results in graphs with exponents of around 1 or less (Fig. 3C), although it is important to not that, in cases where the exponent is close to (or smaller than) 1, the power-law fit becomes somewhat less statistically robust. For this particular starting structure exponents of 1.6 are not observed until *m* is set to 10 (Fig. 3D), indicating that a significantly greater degree of sequence divergence is thus required to recapitulate the degree distribution (and structural diversity) of the PDUG from that region of sequence-structure space. When the folding criterion is relaxed, however, we find that the behavior of the model from both starting structures is similar (see Fig. 2D), implying that it is the nature of "foldability" or designability (14) in the vicinity of this starting structure that gives rise to the difference between runs based on this starting structure compared to the first.

Although degree distributions with $\gamma \sim 1.6$ are readily observed with $m = 8$ and 10 for the two starting structures discussed above, the statistical features of the resulting graphs differ quite significantly from run to run. Graphs with exponents ranging from $-0.8$ to $-1.8$ can be observed in simulations based on the same starting structure and the same evolutionary parameters (see Fig. 3D), indicating that the evolution in this case represents a highly non-ergodic and non-equilibrium sampling of sequence-structure space. Stochastic events early in the simulation seem to set the overall behavior of the graph that is eventually produced by the algorithm, indicating that this model is highly sensitive to random fluctuations especially during the early stages of the evolution. Similar "giant fluctuations" have been observed in other duplication-and-divergence models, such as models describing the evolution of protein-protein interactions (7). Given that the PDUG represents the only available "run" of actual protein evolution, it is unclear the extent to which this heterogeneity might have influenced the degree distribution of the PDUG. We leave further exploration of the relationship between sequence-structure landscape and evolutionary algorithms to future work.

**Clustering Coefficient Distributions**

Although the correspondence between the degree distributions of sequence-based model graphs and that of the PDUG is quite suggestive, the degree distribution represents only one of the statistical features of the network, and one may ask how other features compare between the model graphs and the PDUG. One such feature is the distribution of the clustering coefficient of each node, which is a measure of how many connections exist between a given node's structural neighbors. $C_i(k)$ is the clustering coefficient of node $i$ and is defined as follows (5):

$$C_i(k) = \frac{E_{N,i}}{\frac{k(k-1)}{2}}$$

where $E_{N,i}$ is the number of edges between the neighbors of node $i$ and $k$ is the degree of node $i$. The distribution of $C(k)$ for the PDUG (Fig. 4A) is relatively flat. This distribution is markedly different from that observed for the entire graph of unique 3x3x3 lattice structures (15) (Fig. 4B), which provides both a "random-graph" and polymer control for the $C(k)$ behavior. In order to determine if this distribution is simply a consequence of the scale-free nature of the PDUG, we create a "randomly rewired" version of the PDUG. This randomly rewired graph is created using an algorithm similar to that used in (21): at each step, two edges on the graph are "swapped" such that the degree of each node is maintained. The $C(k)$ distribution for the randomly rewired graph is also markedly different from that of the PDUG (Fig. 4C). This difference is most readily apparent at larger values of $C(k)$—the PDUG contains a preponderance of highly interconnected, "cliquish" regions compared to both the LSG and to the randomly rewired control. We find that the graphs produced by our sequence-based evolutionary model also exhibit relatively flat $C(k)$ distributions with strong bias towards larger values of $C(k)$ (Fig. 4D) that is not observed in randomly rewired model graphs. Very similar C(k) distributions are obtained for graphs obtained from the original nodes-and-edges evolutionary model for the PDUG (2) and non-sequence-based divergent models
1111

sampling from the LSG (15) (data not shown). The sequence-based model discussed above (like other models of the evolution of the PDUG) is thus not only able to recapitulate the degree distribution of the PDUG but other statistical features of the graph as well.

**Discussion**

The results described above represent, to our knowledge, the first demonstration that a scale-free network that describes a particular biological system (such as the PDUG) may be recapitulated using an evolutionary algorithm that attempts to accurately model the underlying biology and physics of the evolution of that system. This constitutes an important extension of graph-theoretic models beyond algorithms in which edges are placed between newly evolved nodes and the rest of the graph according to evolutionarily reasonable but nonetheless highly abstract and artificial rules. Despite this fundamental advancement, this work is nonetheless still a proof of the principle that sequences can divergently sample structural spaces in such a way that scale-free networks similar to the PDUG are produced. Indeed, the large mutational "steps" required to obtain realistic structural diversity in the above model (i.e. $m$ values of 8 or 10) have no clear analogue for real proteins, and reasonable mechanisms underlying large sequence divergence, such as recombination or insertion-deletion, must be implemented in order to develop more accurate models. The realism of the current model, however, is in some ways most severely limited by fact that the "proteins" we consider are constrained to a lattice space, and it is quite unclear how mechanisms such as recombination might be "accurately" built into such a model. We thus leave more realistic physical and mutational models to future work.

Our findings have important implications not only for the study of protein evolution, but also for the evolution of biological network and the study of protein folding. In case of the former, the existence of a successful *a priori* model for the diveregent evolution of protein structures indicates that future models of other biological networks could be based on models of the underlying mechanisms. Indeed, given that many functional features of proteins are in large measure dictated by their structure, one might imagine that the divergent evolution of protein structures has played a dominant role in the evolution of scale-free transcriptional, metabolic and protein-protein interaction networks. Also, the highly non-equilibrium nature of this model has strong implications for the study of protein folding, particularly for the development of residue-residue or atom-atom interaction potentials. It is possible that the nature of sequence-structure sampling over the course of structural evolution may lead to biases in the resulting database of structures that might reduce the accuracy of knowledge-based potentials derived form such databases. To test this hypothesis, one may employ sets of evolved lattice structures to derive knowledge-based potentials and test the resulting potential against the potential used to design (or in this case evolve) the lattice structures (10, 22-24). Such a study would not only provide some indication of the extent to which the highly non-equilibrium nature of structural evolution might have an influence on such potentials but might also lead to the development of more accurate knowledge-based methods.

**Acknowledgements**

We would like to thank Dr. N. Dokholyan, Dr. A. Lesk and Dr. M. Vendrusculo for stimulating discussions. We thank the NIH for financial support. E.J.D. also acknowledges the support of Howard Hughes Medical Institute predoctoral fellowship.

**References**

1. Koonin, E. V., Wolf, Y. I. & Karev, G. P. (2002) *Nature* **420,** 218-23.
2. Dokholyan, N. V., Shakhnovich, B. & Shakhnovich, E. I. (2002) *Proc Natl Acad Sci U S A* **99,** 14132-6.
3. Qian, J., Luscombe, N. M. & Gerstein, M. (2001) *J Mol Biol* **313,** 673-81.
4. Deeds, E. J., Shakhnovich, B. & Shakhnovich, E. I. (2004) *J Mol Biol* **336,** 695-706.
5. Albert, R. & Barabasi, A.-L. (2002) *Rev Mod Phys* **74,** 47-97.
6. Barabasi, A. L. & Albert, R. (1999) *Science* **286,** 509-12.
7. Kim, J., Krapivsky, P. L., Kahng, B. & Redner, S. (2002) *Phys Rev E Stat Nonlin Soft Matter Phys* **66,** 055101.
8. Lesk, A. M., Lo Conte, L. & Hubbard, T. J. (2001) *Proteins* **Suppl 5,** 98-118.
9. Shakhnovich, E. I. & Gutin, A. (1990) *J Chem Phys* **93,** 5967-5971.
10. Mirny, L. A. & Shakhnovich, E. I. (1996) *J Mol Biol* **264,** 1164-79.
11. Tiana, G., Broglia, R. A. & Shakhnovich, E. I. (2000) *Proteins* **39,** 244-51.
12. Li, J., Wang, J., Zhang, J. & Wang, W. (2004) *J Chem Phys* **120,** 6274-6287.
13. Li, H., Helling, R., Tang, C. & Wingreen, N. (1996) *Science* **273,** 666-9.
14. England, J. L. & Shakhnovich, E. I. (2003) *Phys Rev Lett* **90,** 218101.
15. Deeds, E. J., Dokholyan, N. V. & Shakhnovich, E. I. (2003) *Biophys J* **85,** 2962-72.
16. Tiana, G., Shakhnovich, B. E., Dokholyan, N. V. & Shakhnovich, E. I. (2004) *Proc Natl Acad Sci U S A* **101,** 2846-51.
17. Dinner, A. R., Abkevich, V., Shakhnovich, E. & Karplus, M. (1999) *Proteins* **35,** 34-40.
18. Miyazawa, S. & Jernigan, R. L. (1985) *Macromolecules* **18,** 534-552.
19. Miyazawa, S. & Jernigan, R. L. (1996) *J Mol Biol* **256,** 623-44.
20. Goldstein, R. A., Luthey-Schulten, Z. A. & Wolynes, P. G. (1992) *Proc Natl Acad Sci U S A* **89,** 4918-22.
21. Maslov, S. & Sneppen, K. (2002) *Science* **296,** 910-3.
22. Thomas, P. D. & Dill, K. A. (1996) *J Mol Biol* **257,** 457-69.
23. Zhang, L. & Skolnick, J. (1998) *Protein Sci* **7,** 112-22.
24. Chiu, T. L. & Goldstein, R. A. (2000) *Proteins* **41,** 157-63.

**Figure Legends**

**Fig. 1.** Diagram of the structural evolution model. At each step, one existing sequence-structure pair is chosen for duplication. A set of *m* mutations is made to one of resulting duplicates (the other is preserved on the graph unchanged). The native state of the new sequence is the maximally compact lattice structure in which the new sequence exhibits the lowest energy. Folding of the new sequence into this structure is tested via a Z-score procedure as described in the text. If the new sequence folds into its native structure, it is added to the graph, and if not, that sequence structure pair is discarded.

**Fig. 2** Degree distributions for graphs evolved without folding constraints. **A.** The degree distribution of a set of 3500 structures evolved with no folding constraints and *m* set to 1. In this plot, as with all other figures of degree distributions in this work, the degree of each node is increased by 1 in order to allow for display of nodes with degree 0 on log-log plots. Also, the red line in this and all other degree distribution plots in this figure represents a power-law fit of the data, and the indicated exponent is taken from that fit. **B.** The degree distribution for a graph evolved with *m* = 2. **C.** The degree distribution for a graph evolved with *m* = 3. **D.** The degree distribution for a graph evolved from a different starting structure than that employed for **A**, **B** and **C**. In this case, *m* is set to 2.

**Fig. 3** Degree distributions for graphs evolved under the folding criterion. **A.** The degree distribution of a set of 3500 structures evolved with a folding Z-score cutoff of –6 and *m* = 2. The red line in this and all other plots in this figure represents a power-law fit of the data, and the indicated exponent is taken from that fit. **B.** The degree distribution of a graph evolved with a folding Z-score cutoff of –6 and *m* = 8. **C.** The degree

distribution of a graph evolved with a different starting structure than **A** and **B** (the same alternative starting structure employed for Fig. 2D). In this case the folding Z-score cutoff is again set to −6 and *m* is set to 8. **D.** The degree distribution of a graph evolved with the same starting structure used in C, but with *m* = 10.

**Fig. 4** Clustering coefficient distributions. **A.** The distribution of the clustering coefficients of domains on the PDUG. **B.** The distribution of the clustering coefficients for structures in the LSG, which is comprised of all maximally compact 27-mer structures. **C.** The distribution of clustering coefficients for a randomly rewired version of the PDUG. **D.** Comparison of the clustering coefficient distribution between the PDUG, a randomly rewired version of the PDUG and a graph evolved using the evolutionary model. The evolutionary graph here starts with the first seed sequence-structure pair, is evolved using a folding Z-score cutoff of −6 and m = 8, and has a degree distribution with γ ~ 1.6.

**Figure 1**

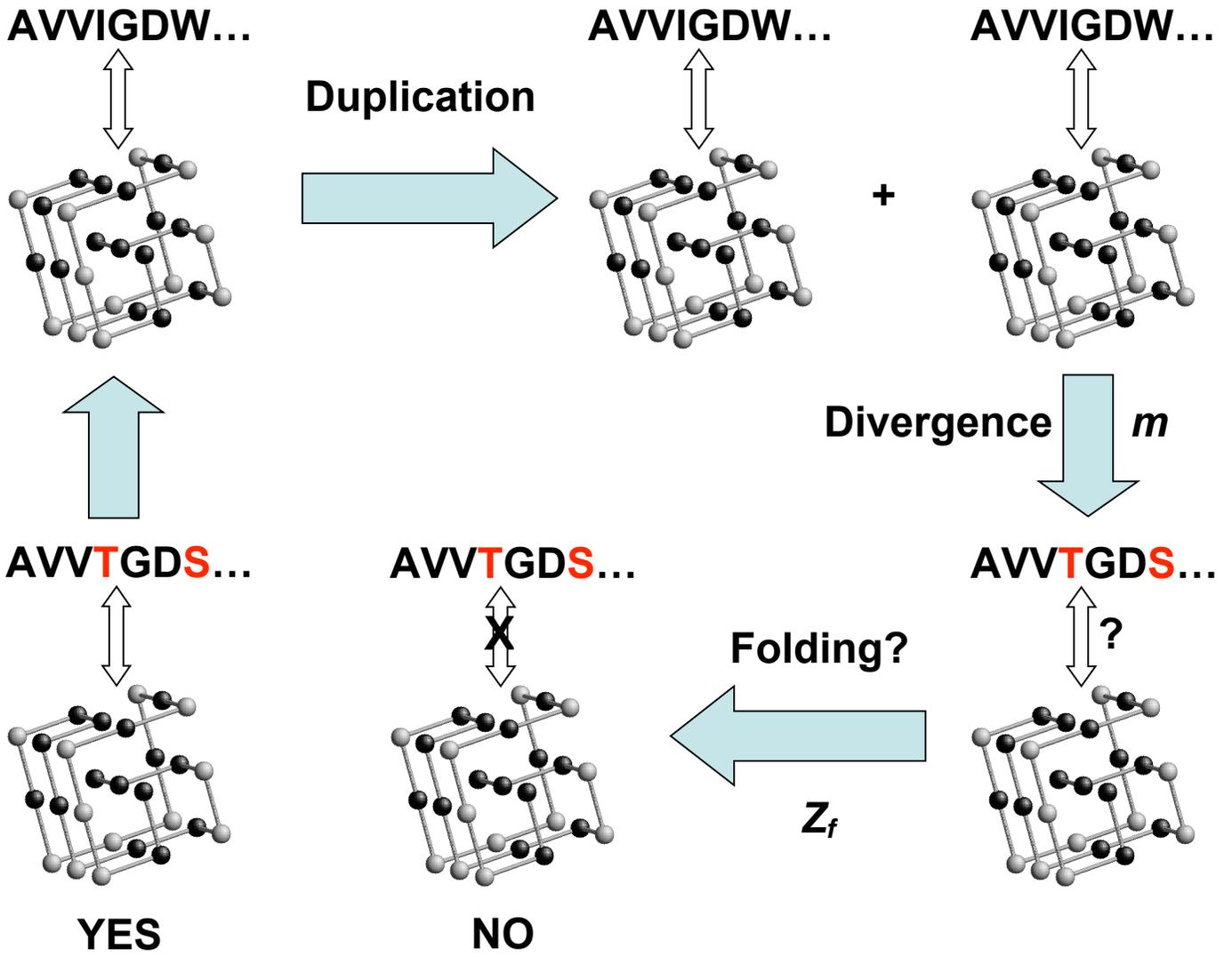

**Figure 2**

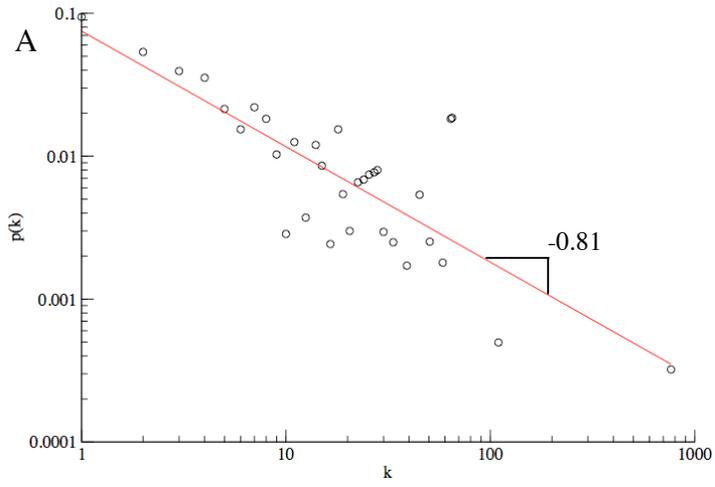
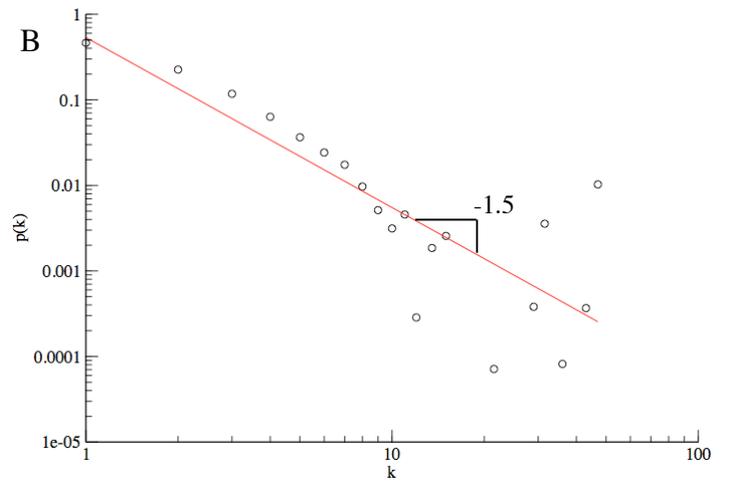
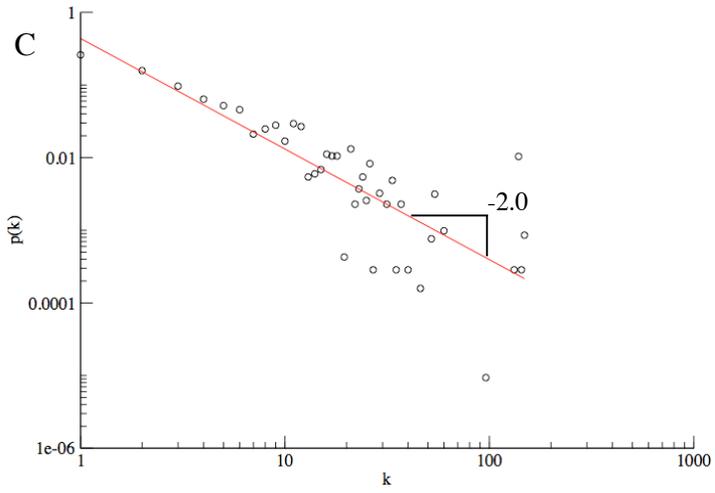
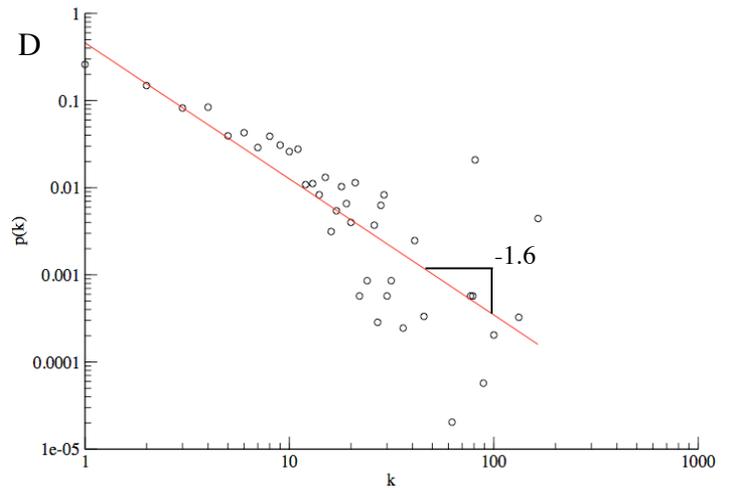

**Figure 3**

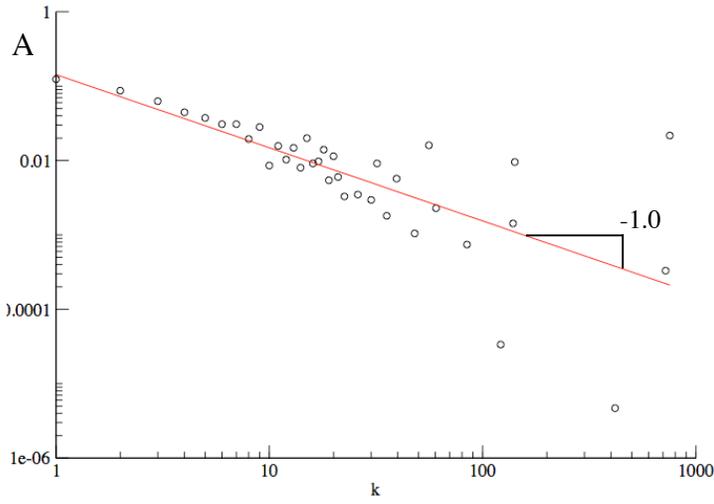
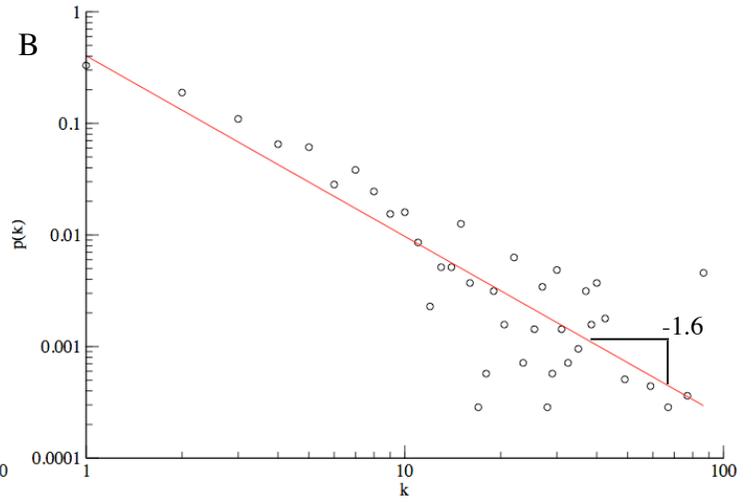
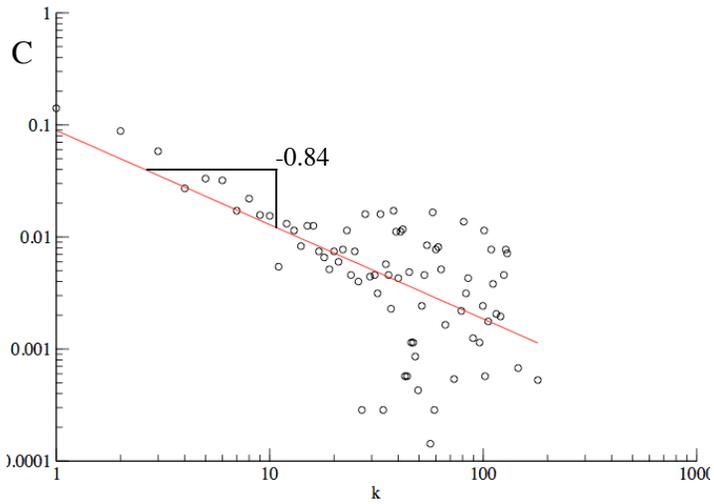
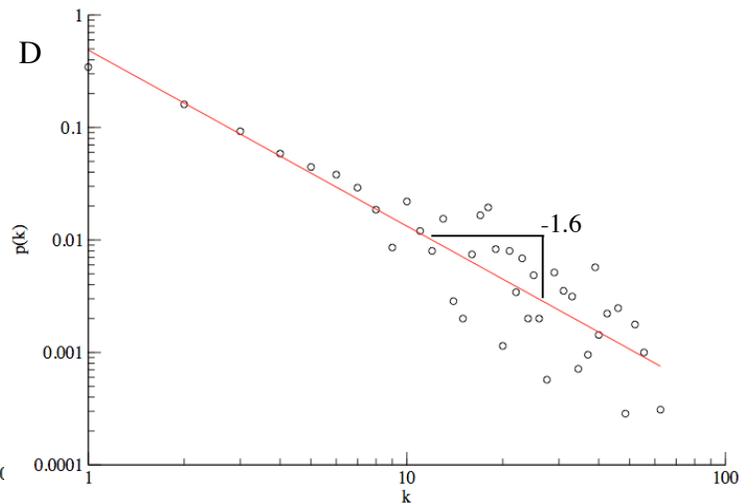

**Figure 4**

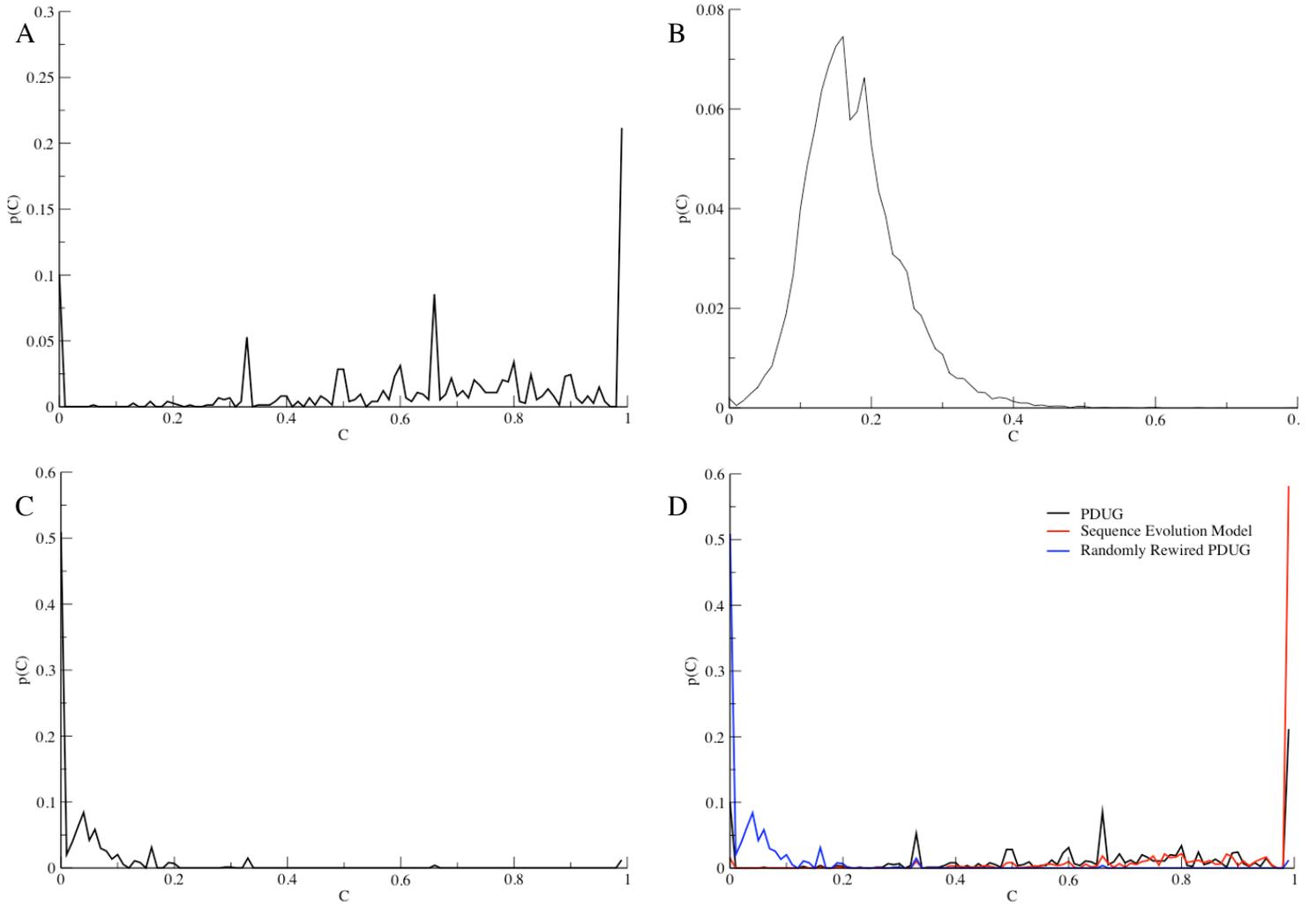